\documentclass[a4]{article}
\usepackage[left=2cm,right=2cm,top=2cm,bottom=2cm,bindingoffset=0cm]{geometry}
\usepackage{epsfig}
\usepackage{epstopdf}
\usepackage{multirow}
\newcommand{\Ref}[1]{(\ref{#1})}
\newcommand{\beao}{\begin{eqnarray*}}
\newcommand{\eeao}{\end{eqnarray*}}
\newcommand{\be}{\begin{equation}}
\newcommand{\ee}{\end{equation}}
\newcommand{\bea}{\begin{eqnarray}}
\newcommand{\eea}{\end{eqnarray}}
\newcommand{\beq}{\begin{eqnarray}}
\newcommand{\eeq}{\end{eqnarray}}
\newcommand{\nn}{\nonumber}

\newcommand{\la}{\lambda}

\begin{document}
\title{ On magnetization of quark-gluon plasma at the LHC experiment energies}
\author{
V.~ Skalozub\thanks{e-mail: Skalozubv@daad-alumni.de}\\
{\small Oles Honchar Dnipropetrovsk National University, 49010 Dnipro, Ukraine}\\
\\
P.~ Minaiev\thanks{e-mail: Minaevp9595@gmail.com}\\
{\small Oles Honchar Dnipropetrovsk National University, 49010
Dnipro, Ukraine}}

\date{}
\maketitle

\begin{abstract}
Large scale chromomagnetic,  $B_3, B_8$, and usual magnetic,$H$,
fields have to be generated in $QCD$ after the deconfinement phase
transition ($DPT$) at temperatures $T $ larger than deconfinement
temperature $T_d$. The two former  fields are created
spontaneously due to asymptotic freedom of gluon intaractions.
Whereas $H$ is produced  due to either the feature of quarks to
possess both electric and color charges or a vacuum polarization
in this case. At the polarization, the vacuum quark loops mix the
external fields. As a result, $B_3, B_8$ become the sources
generating $H$. The latter field appears at $T$ much lower than
the electroweak phase transition temperature $T_{ew}$. This
mechanism should exhibit itself  at the LHC experiments on heavy
ion collisions. It operates at the one-loop diagram level for an
effective potential. The created fields are temperature dependent
and occupying the macroscopic volume of quark-gluon plasma.  The
magnetization influences different processes and may serve as a
signal for the $DPT$.
\end{abstract}

\section{Introduction}
Studying of $QCD$ vacuum, its dynamics and properties, is one of
the most important problems investigated at the LHC experiments.
Here, in heavy ion collisions a new matter phase - quark-gluon
plasma ($QGP$)- is expected to be produced. It consists of quarks
and gluons deliberated from hadrons
 at high temperature $T > T_d \sim $ 180 - 200 MeV, where $T_d$ is
 a deconfinement temperature. In theory, investigation of this phase transition and $QGP$ properties
was carried out by different method - analytic perturbative and
nonperturbative, various numerical methods and Monte-Carlo
simulations on a lattice. Details on these studies and results
have been presented in numerous publications (see, for example
\cite{Satz2012} -  \cite{szab14-LATTICE2013-014}). In particular,
$QGP$ had been existed in the hot Universe and influenced various
processes \cite{gras00-348-163}, \cite{eliz12-72-1968}. One of the
distinguishable properties of gluon fields at high temperature is
a spontaneous vacuum magnetization closely related with asymptotic
freedom. It happen due to a large magnetic moment of  charged
color gluons (gyromagnetic ratio $\gamma = 2$) and results in a
stable, temperature dependent classical chromo(magnetic)
 fields occupying large domains of space.
 The magnetization has been investigated in detail in $SU(3)$ gluodynamics
\cite{Strelchenko2004} by analytic methods and in $SU(2)$
gluodynamics \cite{Demchik2008}, \cite{Antropov2014} by the
Monte-Carlo simulations on a lattice. In both cases, the creation
of the fields  was detected.

In the present paper, we investigate a scenario corresponding to
experiments on heavy ion collisions at the LHC. It  is realized in
$QGP$ at temperatures $T > T_d$ and related with quarks. The
effect consists in generation of  usual magnetic field $H$ due to
the vacuum polarization of quark fields by the constant color
magnetic fields $B_3$ and $B_8$, which have to be created
spontaneously in the gluon sector after the $DPT$. Really, quarks
possess both electric $e$ and color $g$ charges. Therefore, due to
vacuum polarization, the  mixing of external fields happens and
the corresponding  terms in an effective potential  may serve as
specific sources for $H$. This mechanism has to start to operate
at temperatures of $QGP$ $T_d < T << T_{ew}$, where $T_{ew}\sim
100$ GeV is an electroweak phase transition temperature. At these
$T$ the non-Abelian $SU(2)$ constituent  of the electromagnetic
field is screened by the scalar condensate. So, the field $H$
could not be generated spontaneously in the electroweak sector of
the standard model \cite{Demchik2015}. It is expected that the
magnetic field $H$ is temperature dependent and occupying a large
plasma volume as $B_3$ and $B_8$ do. According to this scenario,
the color magnetic fields, generated spontaneously after the
$DPT$, play the role of the specific sources for usual
 magnetic field.

To investigate this possibility, we calculate the one-loop
contribution to the effective potential  $V^{(1)}(B_3,B_8,H,T)$ of
quark fields in the external color magnetic fields $B_3$ and $B_8$
and usual magnetic field $H$ at finite temperature and investigate
its properties for  $T > T_d$. Two specific approximations will be
applied. First is high temperature expansion for the effective
potential derived by means of the Mellin transformation. Second is
a low temperature expansion which also is relevant. It this way
the effective potentials of the external fields generated by the
quark loop are derived. Assuming that the fields $B_3(T)$ and
$B_8(T)$ appear spontaneously after the $DPT$ due to vacuum
polarization of gluon fields, as it was derived in
\cite{Strelchenko2004}, and taking into consideration the specific
values of these fields corresponding to temperatures not much
higher of $T_d$, we estimate the strength of the magnetic field
$H$.

The paper is organized as follows. In sect. 2 the one-loop
effective potential $V^{(1)}(B_3,B_8,H,T)$  of external fields and
temperature generated by quark contributions is calculated. In
sect. 3 the high temperature and low temperature asymptotic
expansions are derived. Then, in sect. 4 the generated magnetic
field $H$  is investigated. Discussion and conclusions are given
in the last section.

\section{Effective potential of external fields}
In what follows, we consider the situation when the temperature of
the $QGP$ is not much higher than $T_d$. So, according to
\cite{Strelchenko2004}, both color magnetic fields $B_3$ and $B_8$
are spontaneously created in the gluon sector of $QCD$. They will
be considered as the background for calculating  the one-loop
quark effective potential $V^{(1)}(B_3,B_8,H,T)$.

 To be in correspondence with
the notation and results of this paper, we present the $SU(3)_c$
gluon field in the form
\be \label{field} A_\mu^a = B_\mu^a + Q_\mu^a, \ee
where $B_\mu^a$ is background classical field and $Q_\mu^a$
presents quantum gluons. We choice the external field potential in
the form $B_\mu^a = \delta^{a3} B_{3\mu} + \delta^{a8} B_{8\mu}$,
where $B_{3\mu} = H_3 \delta_{\mu  2} x_1$ and $B_{8\mu} = H_8
\delta_{\mu  2} x_1$  describe constant chromomagnetic fields
directed along third axis in the Euclidean space and $a = 3$ and
$a = 8$ in the color $SU(3)_c $ space, respectively. The field
tensor has components: $F^{ext ~a}_{\mu\nu}= \delta^{a3}
F_{3\mu\nu} + \delta^{a8} F_{8\mu\nu}, F_{c12} = - F_{c21} = H_c,
 c = 3, 8 $. Accounting for this choice, we direct usual magnetic
 field also along third axis and write the electromagnetic potential in the similar
 form: $A_{\mu}^{ext} = H \delta_{\mu 2} x_1$.

In \cite{Strelchenko2004} it was obtained that at temperatures $T
\geq T_d$ both the fields
\bea \label{H3,8} g H_3(T) &=& 0.2976 \frac{g^4 T^2}{\pi^2},\\ \nn
 g H_8(T) &=& 0.9989 \frac{g^4 T^2}{\pi^2}\eea
are spontaneously generated. This situation is changed at
asymptotically high temperatures when only the field $H_3$ is
created. Below, we consider the first more general case and
concentrate on the qualitative picture of the phenomenon. We take
into consideration these fields as external ones without
accounting of the quark corrections to the values \Ref{H3,8}. They
are numerically small and do not change significantly final
results.

To consider the generation of $H$ by the fields $H_3, H_8$ we
first calculate the quark spectrum in the presence of all these
fields. The corresponding Dirac equation reads
\be \label{Deq} (i \gamma_\mu D_\mu + m_f ) \psi^a = 0, \ee
where $\psi^a$ is a quark wave function, $a$ is color index, $m_f$
is mass of $f$-flavor  quark.  The covariant derivative describes
the interactions with external magnetic fields $H$ and $H_3, H_8$:
\be \label{D} D_\mu = \partial_\mu - i e A_\mu^{ext}  - i g ( T^3
B_\mu^3 + T^8 B_\mu^8 ) \ee
where $T^3 = \frac{\lambda^3}{2}, T^8 =\frac{\lambda^8}{2}$ are
the generators of $SU(3)$ group, $\la^{3,8}$ are Gell-Mann
matrixes. Due to the choice of the potentials and the Abelian
nature of the fields we can  present the quark spectrum as the sum
of contributions of the three following external fields:
\bea \label{fields} q H_1 &=& q_f H + \frac{g}{2} (H_3 +
\frac{1}{\sqrt{3}}H_8), \\ \nn q H_2 &=& q_f H + \frac{g}{2} (-
H_3 + \frac{1}{\sqrt{3}}H_8 ) , \\ \nn q H_3 &=& q_f H -
g\frac{1}{\sqrt{3}}H_8. \eea
Here, $q_f$ is electric charge of   $f$-quark. Each flavor energy
spectrum is given by the  known expression (see, for instant,
\cite{Akhiezer69} )
\be \label{spectrum} \epsilon^2_{i,n,\rho,f} = m^2_f + p^2_z + (2
n + 1) q H_i - \rho   q H_i, \ee
where $p_z$ is momentum along the field direction,  $\rho = \pm
1.$

 Vacuum energy is defined as the sum of the modes having
negative energy, which  also is  well known \cite{Akhiezer69}
\be \label{vac0} V^{(0)}_{l = 0} = \frac{1}{8 \pi^2} \sum_{f =
1}^6 \sum^3_{i = 1} \int\limits^{\infty}_0 \frac{ d ~s}{s^3} e^{-
m^2_f s} \bigl[q H_i s \coth (q H_i s) - 1 \bigr].\ee
Next step is to account for finite temperature. In the imaginary
time formalism for fermions, it is reduced to the summation over
discrete odd   imaginary energy $p_4 = \frac{(2l + 1) \pi}
{\beta}$, $\beta = 1/T$ is inverse temperature
\cite{kala84-32-525}, \cite{Satz2012}. The result yields
\be \label{VT} V =  \frac{1}{8 \pi^2} \sum_{f = 1}^6 \sum^3_{i =
1}\sum^{\infty}_{l = - \infty}(- 1)^l \int\limits^{\infty}_0
\frac{ d ~s}{s^3} \exp (- m^2_f s - \frac{\beta^2 l^2}{4 s})
\bigl[q H_i s \coth (q H_i s) - 1 \bigr].\ee
This is expression of interest. It will be investigated in the
next section. Note, the term with $l = 0$ is the vacuum energy
\Ref{vac0}.
\section{Asymptotic expansion}
The expression \Ref{VT} includes  complete information about the
 polarization. In particular,
it describes the nonlinear effective potentials of field
interactions. To derive them at finite temperature we make a high
temperature and low temperature expansion of \Ref{VT}. In the
first case, we apply Millin's transformation for summing up the
series in \Ref{VT}. This method is described in detail in
\cite{Weldon1982}. In what follows,  we use the dimensionless
variables and measure all the parameters in units of proton mass
$m_p$. The dimensionless constituent quark mass is $m_f =
m_f/m_p$, the field strengths $h_i = (q H_i)/m_p^2,$ and the
temperature $T = T/m_p$. In these units the dimensionless
potential \Ref{VT} is $V/m_p^4$.

Expanding $\coth (q H_i s) $ in series, performing integration
over $s$ and applying Mellin's transformation for the obtained
expressions we find for the temperature dependent part in the high
temperature limit ( $T \to \infty$)
\bea \label{highT}  12\pi^2 V^{(T)} &=& \frac{2}{3}\sum_i
h_i^2\left[
\frac{1}{2}\left(\gamma+ln(\frac{\omega}{\pi})\right)+\frac{7\zeta'(-2)}{4}\omega^2+\frac{31\zeta'(-4)}{64}\omega^4
\right] \\ \nn & -&\frac{1 }{90\mu^4}\sum_i h_i^4\left[
-1+\frac{31\zeta'(-4)}{8}\omega^4 \right].\eea
In the low temperature limit $(T \to 0$ )  we obtain
\bea \label{lowT}  12\pi^2 V^{(T)} &=& -\frac{2}{3} \sum_i
h_i^2\sqrt{\frac{\pi}{2}}e^{-\omega}\left[
\frac{1}{\sqrt{\omega}}-\frac{1}{8\omega\sqrt{\omega}}\right] \\
\nn &+& \frac{1}{90\mu^4}\sum_i h_i^4
\sqrt{\frac{\pi}{2}}e^{-\omega}\left[
\omega\sqrt{\omega}+\frac{15}{8}\sqrt{\omega}
+\sqrt{\frac{1}{\omega}}\frac{105}{128}\right], \eea
where $h_i$ are dimensionless fields \Ref{fields}.
 We also introduced
the notations: \be x_3=\frac{B_3 Q_c}{m_p^2},\quad x_8=\frac{B_8
Q_c}{m_p^2},\quad x=\frac{H e}{m_p^2},\ee
\be \quad\omega=\beta m_f=\mu m_p\beta,\quad \mu=\frac{m_f}{m_p}.
\ee
 From \Ref{highT}, \Ref{lowT} we see that different kind  mixing of fields takes place.
In particular, there are the terms of the form
\be \label{moments} \sim eH\times  (g H_3)^3, \sim eH\times (g
H_8)^3, ...,\ee
which mean that color fields $H_3, H_8$ play the role of external
sources (magnetic moments, etc.) for usual magnetic field $H$ and
generate it. To verify this idea, we have to consider the
tree-level plus $V $ effective potentials $V_{eff.} =
\frac{1}{2}H^2 + V(H, H_3, H_8, T)$ and solve the stationary
equation
\be \label{fieldH} \frac{\partial V_{eff.}}{\partial H}  = 0. \ee
If it has the nonzero solutions $H = H_c$   and  the energy of
them is negative, we determine the magnetic field produced after
the DPT. Investigation of this will be done in the next section.
\section{Generation of magnetic field}
Let us present the solutions to equation  \Ref{fieldH} in high
temperature \Ref{highT} and low temperature \Ref{lowT}
approximations to the effective potential. From the point of view
of applications, the former case is more adequate to  $QGP$ at the
LHC experiments.

In Table 1  we show  the values of $H_3, H_8$ obtained according
to formulae \Ref{H3,8} and $H$ generated according to \Ref{fieldH}
at temperatures $ T = 180, 200, 220$ MeV corresponding to $T \geq
T_d$. The case of  only one constituent quark mass $m_u$ is
investigated. The values of the magnetic field energy are also
adduced.
 In units of $m_p$ and
assuming for estimates that $\alpha_s = \frac{g^2}{4 \pi} = 1 $
the color magnetic field strengths  read
\be \label{fields3,8T}  x_3 = \frac{g H_3}{m_p^2} =
0.76164\frac{T^2}{m_p^2},~~x_8 = \frac{g H_8}{m_p^2} = 15.9824
\frac{T^2}{m_p^2}. \ee
%

%
%
\begin{center}
\begin{table}[h]
\begin{tabular}{|c|c|c|c|c|c|c|}
\hline
T, MeV&$x_8$&$x_3$&$V_{eff}^{(h)} 10^{-2}$&$x^{(h)} 10^{-3}$&$V_{eff}^{(l)} 10^{-2}$&$x^{l}10^{-3}$\\
\hline
180&$0.59$&$0.18$&$-0.153$&$-0.546$&$-0.150$&$-0.531$\\
\hline
200&$0.73$&$0.22$&$-0.211$&$-0.575$&$-0.202$&$-0.546$\\
\hline
220&$0.88$&$0.26$&$-0.272$&$-0.582$&$-0.251$&$-0.528$\\
\hline

\end{tabular}

\end{table}

Table 1.The strength values of the fields generated at the typical
temperatures
\end{center}

The solutions for considered  asymptotic expressions  are close to
each other. So,  both these expansions are equally applicable.
 The energy of magnetic
field  is negative decreasing function of $T$.   The total energy
of fields $H_3, H_8$ and $H$ is also negative. So, these field
configurations have to be energetically favorable. As we see, the
field strength values $h$ are two order smaller and orientation is
opposite to the direction of the color field ones.

 As a result, all this is signaling the
generation of macroscopic magnetic field $H$ in the presence of
color magnetic fields $H_3, H_8$, which act as the sources for it.
The field is occupying all the $QGP$ volume where the color fields
present. Of course, these are estimates. More detailed analysis
has to account for all the quark flavors. But in this paper we
concentrate on the qualitative picture of the phenomenon. We
expect that only quantitative changes could happen.
\section{Discussion and conclusion}
As it follows from the obtained results, at temperatures $T > T_d$
in $QGP$ either large scale Abelian chromomagnetic $H_3, H_8$ or
usual magnetic $H$ fields have to present. The mechanisms for
generation of them are quite different. The color fields are
generated at high temperature in the $SU(3)_c$ sector of the
standard model due to gluon vacuum polarization
\cite{Strelchenko2004}, \cite{Antropov2014}. They exist in space
till the color is screened at low temperature. Formally, such type
fields are solutions to field equations without sources. The
$SU(2)$ component of usual magnetic field is also produced
spontaneously at temperatures $T$
 larger than electroweak phase transition temperature $T_{ew} \sim 100$
 GeV due to $W$-boson vacuum polarization.
 But it is screened by  the
 scalar field condensate appeared after this phase transition
 \cite{Demchik2015}.

With temperature lowering, in the interval $T_{ew} > T
> T_d$, when color magnetic fields are present, the field
$H$ esquires other mechanism for generation. It is induced by the
vacuum polarization of quark fields, as it was shown above. Due to
nonlinearity of the polarization, the color fields $H_3, H_8$
become the sources producing $H$.   All these fields occupy
macroscopic volumes of $QGP$. One of consequences of the
magnetization is the discrete spectra of color and/or electric
charged particles, as it is shown for quarks in \Ref{spectrum}. In
fact, this is the distinguishable feature of $QGP$. At low
temperatures, after the confinement the macroscopic magnetic
fields are screened. Only the fields produced by charged currents
remain.

As corollary, we have derived that at the $LHC$ experiment
energies the $QGP$ has to be magnetized, that may serve as signal
for the $DPT$.  To give concrete numerical  estimates one has to
account for other quark flavors and long range correlation
corrections. All these contributions to the effective potential as
well as specific processes signaling the plasma magnetization will
be reported elsewhere.
%
%


\begin{thebibliography}{10}

\bibitem{Satz2012}
Helmut Satz.
\newblock {\em Extreme States of Matter in Strong Interaction physics}, volume
  841 of {\em Lecture Notes in Physics}.
\newblock Springer, 2012.

\bibitem{Greensite2011}
Jeff Greensite.
\newblock {\em An Introduction to the Confinement Problem}, volume 821 of {\em
  Lecture Notes in Physics}.
\newblock Springer, 2011.

\bibitem{kala84-32-525}
O.~K. Kalashnikov.
\newblock {QCD at finite temperature}.
\newblock {\em Fortsch. Phys.}, 32:525, 1984.

\bibitem{bali12-CX-197}
G.~S. Bali et~al.
\newblock {Thermodynamic properties of QCD in external magnetic fields}.
\newblock {\em PoS}, ConfinementX:197, 2012.

\bibitem{levk14-112-012002}
L.~Levkova and C.~DeTar.
\newblock {Quark-gluon plasma in an external magnetic field}.
\newblock {\em Phys.~Rev.~Lett.}, 112(1):012002, 2014.
\bibitem{szab14-LATTICE2013-014}
Kalman Szabo.
\newblock {QCD at non-zero temperature and magnetic field}.
\newblock {\em PoS}, LATTICE2013:014, 2014.



\bibitem{bali14--182}
G.~S. Bali, F.~Bruckmann, G.~Endr{\"o}di, and A.~Sch{\"a}fer.
\newblock {Magnetization and pressures at nonzero magnetic fields in QCD}.
\newblock {\em PoS}, LATTICE2013:182, 2014.
\bibitem{gras00-348-163}

Dario Grasso and Hector~R. Rubinstein.
\newblock {Magnetic fields in the early universe}.
\newblock {\em Phys. Rept.}, 348:163--266, 2001.

\bibitem{eliz12-72-1968}
E.~Elizalde and V.~Skalozub.
\newblock {Spontaneous magnetization of the vacuum and the strength of the
  magnetic field in the hot Universe}.
\newblock {\em Eur. Phys. J.}, C72:1968, 2012.

\bibitem{Strelchenko2004}
V.V. Skalozub and A.V. Strelchenko.
\newblock {On the generation of Abelian magnetic fields in $SU(3)$
gluodynamics at high temperature}.
\newblock {\em Eur. Phys. J. C }, 33: 105, 2004.

\bibitem{Demchik2008}
V.~Demchik and V.~Skalozub.
\newblock{Spontaneous creation of chromomagnetic field  and $A(0)$ condensate at high temperature on a lattice}
\newblock{\em J. Phys. A}, 41: 16405, 2008.

\bibitem{Akhiezer69}
A.I. Akhiezer, V.B. Berestetski.
\newblock{Kvantovaya electtrodinamika}.
\newblock{"Nauka", Moscow}, 624 p., 1969.

\bibitem{Antropov2014}
S.~Antropov, M.~Bordag, V.~Demchik and V.~Skalozub.
\newblock{Long range chromomagnetic fields at high temperature}.
\newblock{Intern. J. Mod. Phys. A}, 26:4831, 2011.

\bibitem{Weldon1982}
H. E.~Haber and H. A. Weldon.
\newblock{On the relativistic Bose-Einstein integrals}.
\newblock{J. Math. Phys.}, 23:1852, 1982.

\bibitem{Demchik2015}
Vadim Demchik and Vladimir Skalozub.
\newblock {Spontaneous magnetization of a vacuum in the hot Universe and
  intergalactic magnetic fields}.
\newblock {\em Phys. Part. Nucl.}, 46(1):1--23, 2015.



\end{thebibliography}
\end{document}